\documentstyle[12pt]{article}
\title{\bf
$U_L(N)\times U_R(N)$-invariant four-fermion interactions and Nambu-Goldstone
mechanism at finite temperature\thanks{This work was supported partially by
National Natural Science Foundation of China and by Grant No.LWTZ-1298 of the
Chinese Academy of Sciences.}}
\author{
{\bf 
Bang-Rong Zhou}\thanks{Regular Associate of The Abdus Salam ICTP.} \\
\normalsize Department of Physics, Graduate School at Beijing \\
\normalsize University of Science and Technology of China,
  Academia Sinica, Beijing 100039, {\bf China}\thanks{
Mailing Address.} \\
\normalsize and  \\
\normalsize CCAST (World Laboratory), Beijing 100080, {\bf China} \\
}
\date {}
\begin{document}
\hoffset = -1 truecm
\voffset = -2 truecm
\baselineskip = 12pt
\maketitle

\begin{abstract}
In a chiral $U_L(N)\times U_R(N)$ fermion model of NJL-form, we prove that,
if all the fermions are assumed to have equal masses and equal chemical
potentials, then at the finite temperature $T$ below the symmetry restoration
temperature $T_c$, there will be $N^2$ massive scalar composite particles and
$N^2$ massless pseudoscalar composite particles (Nambu-Goldstone bosons). This
shows that the Goldstone Theorem at finite temperature for spontaneous symmetry
breaking $U_L(N)\times U_R(N) \to U_{L+R}(N)$ is consistent with the real-time
formalism of thermal field theory in this model.
\end {abstract}
PACS numbers: 11.10.Wx, 11.30.Rd, 11.30.Qc, 14.80.Mz \\
Key words: Fermion condensates, Real-time thermal field theory,
           $U_L(N)\times U_R(N)$ chiral symmetry breaking,
           Nambu-Goldstone bosons

\section{Introduction} 
The Nambu-Goldstone mechanism [1-3] characterizes spontaneous breaking of a
contineous symmetry.  While it has been researched extensively at zero 
temperature field theory, it is still interesting to examine  how this 
mechanism manifests itself at a finite temperature for  a deeper
understanding of symmetry breaking at high temperature, especially of the
consistency between the Nambu-Goldstone mechanisim and the real-time
thermal field theory's formalism [4].  This consistency is not trival in some
models. In this paper, as an example, we will take a simple model of
Nambu-Jona-Lasinio (NJL) -form [2] with $U_L(N)\times U_R(N)$- chirally
-invariant four-fermion interactions to explore this problem.  In the fermion
bubble graph approximation we will calculate the propagators of the scalar and
pseudoscalar bound state modes, determine the masses of these bound states and
finally confirm that, under some conditions, the Nambu-Goldstone mechanism will
be consistent with the real-time formalism of thermal field theory in this
model.  \\
\section{Model}
For a fermion system with $U_L(N)\times U_R(N)$ -invariant four-fermion 
interactions, its Lagrangian can be generally expressed by 
\begin{eqnarray*}
{\cal L}&=&\sum_{j,k=1}^{N}\left[
\bar{Q}_L^j i\not\!{\partial}Q_{Lj}+\bar{Q}_R^j i\not\!{\partial}Q_{Rj}+
g(\bar{Q}_L^jQ_{Rk})(\bar{Q}_R^kQ_{Lj}) \right.\\
& & \left.+g_L(\bar{Q}_L^j\gamma^{\mu}Q_{Lj})(\bar{Q}_L^k\gamma_{\mu}Q_{Lk})
+g_R(\bar{Q}_R^j\gamma^{\mu}Q_{Rj})(\bar{Q}_R^k\gamma_{\mu}Q_{Rk})\right],
\end{eqnarray*} 
$$\eqno(2.1) $$
\noindent where the fermion fields $Q_L$ and $Q_R$ are respectively assigned
in the $N$-dimension representations of the symmetry group $U_L(N)$ and 
$U_R(N)$ and $g$, $g_L$ and $g_R$ are the real coupling constants of the 
corresponding four-fermion interactions. It is indicated that the independent
four-fermion couplings have only the scalar and the vector coupling terms
appearing in Eq.(2.1). By Fierz Rearrangement Theorem, it can be proven that
the tensor coupling with $\sigma^{\mu \nu}$ does not exist and all the other
couplings including the ones with $\gamma_5$, $\gamma_5\gamma_{\mu}$ and the
vector coupling between $Q_L$ and $Q_R$ fields can be transformed into the 
forms shown in Eq.(2.1).  By means of
$$\left.\matrix{
                Q_{Lj}\cr
                Q_{Rj}\cr}
  \right\}=\frac{1}{2}(1\mp\gamma_5)Q_j
\eqno(2.2)$$
\noindent we can rewrite the scalar coupling among the chiral fields by
\begin{eqnarray*}
{\cal L}_{int}^{CS}&=&g\sum_{j,k=1}^{N}(\bar{Q}_L^jQ_{Rk})(\bar{Q}_R^kQ_{Lj})\\
&=&\frac{g}{4}\left[
\sum_{j=1}^{N}{(\bar{Q}^jQ_j)}^2+
\sum_{j\neq k=1}^{N}(\bar{Q}^jQ_k)(\bar{Q}^kQ_j)-
\sum_{j,k=1}^{N}(\bar{Q}^j \gamma_5 Q_k)(\bar{Q}^k \gamma_5 Q_j)\right]. \\
\end{eqnarray*} $$\eqno(2.3)$$
\noindent
Assuming that the scalar couplings $(g/4)\sum_{j=1}^N{(\bar{Q}^jQ_j)}^2$ among
the same $Q_j$ fields could lead to formation of the condensates $\langle
\bar{Q}^jQ_j\rangle \ (j=1,...,N)$ and generation of the fermion masses 
$m_j \ (j=1,...N)$. At finite temperature $T$, the condensate $\langle
\bar{Q}^jQ_j\rangle $ must be replaced by the corresponding thermal expectation
value  ${\langle\bar{Q}^jQ_j\rangle}_T $, thus we will obtain the gap equation
for the dynamical fermion mass at $T\neq 0$
$$m_j(T, \mu_j)=-\frac{g}{2}{\langle\bar{Q}^jQ_j\rangle}_T. \eqno(2.4)$$
\noindent A natural supposition is that
$$\mu_1=\mu_2=\ldots=\mu_N=\mu \ \ {\rm and}\ \ 
  m_1=m_2=\ldots =m_N=m\equiv m(T, \mu), \eqno(2.5) $$
\noindent i.e. the chemical potentials and the dynamical masses of all the 
fermions are equal, then the gap equation will take the form
$$gI=1 \eqno(2.6)$$
\noindent with
\begin{eqnarray*}
I&=&\frac{1}{2m}\int\frac{d^4l}{{(2\pi)}^4}tr[iS^{11}(l,m)]\\
 &=&2\int \frac{d^4l}{{(2\pi)}^4}\left[
\frac{i}{l^2-m^2+i\varepsilon}-2\pi\delta(l^2-m^2){\rm sin}^2\theta(l^0,\mu)
\right], \ \ \ \ \ \ \ \ \ \ \ \ \ \ \ \ \ \ \ \ \ \ \ \ \ \ \ (2.7)
\end{eqnarray*}
\noindent where we have used the fermion matrix propagator in the real-time 
thermal field theory [4]
\begin{eqnarray*}
\lefteqn{\left(\matrix{
         iS^{11}(l,m) & iS^{12}(l,m)\cr
         iS^{21}(l,m) & iS^{22}(l,m)\cr
         }\right)=
  \left(\matrix{
         i/(\not\!{l}-m+i\varepsilon) & 0 \cr
         0 & -i/(\not\!{l}-m-i\varepsilon)\cr
         }\right)}  \\
    & &   -2\pi(\not\!{l}+m)\delta(l^2-m^2)
  \left(\matrix{
         {\rm sin}^2\theta(l^0,\mu) & 
         \frac{1}{2}e^{\beta\mu/2}{\rm sin}2\theta(l^0,\mu) \cr
        -\frac{1}{2}e^{-\beta\mu/2}{\rm sin}2\theta(l^0,\mu) & 
         {\rm sin}^2\theta(l^0,\mu) \cr
         }\right) 
\end{eqnarray*} $$\eqno(2.8)$$
\noindent with $\beta=1/T$ and
$${\rm sin}^2\theta(l^0,\mu)=\frac{\theta(l^0)}{e^{\beta(l^0-\mu)}+1}+
                           \frac{\theta(-l^0)}{e^{\beta(-l^0+\mu)}+1}.
                              \eqno(2.9)$$
\noindent It is pointed out that the gap equation (2.6) could be satisfied
only at the temperature $T<T_c$, where $T_c$ is the critical temperature for
chiral symmetry restoration in a model of NJL-form [5]. in the following
discussions of the propagators of bound states we will confine ourselves to 
the temperature below $T_c$, i.e. assume Eq.(2.6) is satisfied.
\section{Scalar bound state modes}  
\indent The propagators for scalar bound states relate to the scalar 
four-point functions of fermions. To calculate them in the real-time formalism
of thermal field theory, we must take the doubled scalar four-fermion 
interaction Lagrangian [4]
$${\cal L}_{int}^S=\frac{g}{4}\sum_{a=1,2}\sum_{j,k=1}^{N}
{(-1)}^{a+1}{(\bar{Q}^jQ_k)}^{(a)}{(\bar{Q}^kQ_j)}^{(a)}, \eqno(3.1)$$
\noindent where $a=1$ means physical fields and $a=2$ ghost fields.  The 
physical and the ghost fields can interact only through propagators.  Consider
scalar bound state $(\bar{Q}^kQ_j)$.  The scalar four-point function from 
$a$-type vertex to $b$-type vertex will be denoted by 
$$\Gamma_S^{{(\bar{Q}^jQ_k)}^{(b)}{(\bar{Q}^kQ_j)}^{(a)}}(p)\equiv 
\Gamma_S^{ba}(p),$$
\noindent then in the bubble graph approximation, they submit to the
following equations
$$\Gamma_S^{ba}(p)=i\frac{g}{2}{(-1)}^{a+1}\delta^{ba}+
\sum_{c=1,2}\Gamma_S^{bc}(p)L^{ca}(p)i\frac{g}{2}{(-1)}^{a+1}, \ \ a,b=1,2,
\eqno(3.2)$$
\noindent which are extension of the similar equations at zero temperature [6]
to finite temperature, where $p$ is the four-momentum of the bound state
$(\bar{Q}^kQ_j)$, $L^{ca}(p)$ expresses the contribution of the $Q_j-\bar{Q}^k$
fermion loop with an $a$-type and a $c$-type scalar interaction vertex,
$$L^{ca}(p)\equiv L_{\bar{Q}^kQ_j}^{ca}(p)=
-\int \frac{d^4l}{{(2\pi)}^4}tr[iS^{ca}(l,m)iS^{ac}(l+p,m)]. \eqno(3.3)$$
\noindent Eq.(3.2) has the solutions
$$\Gamma_S^{b1}(p)=\frac{1}{\Delta(p)}\left\{
i\frac{g}{2}\left[1+i\frac{g}{2}L^{22}(p)\right]\delta^{b1}+
\frac{g^2}{4}L^{21}(p)\delta^{b2}\right\},$$
$$\Gamma_S^{b2}(p)=\frac{1}{\Delta(p)}\left\{
\frac{g^2}{4}L^{12}(p)\delta^{b1}-
i\frac{g}{2}\left[1-i\frac{g}{2}L^{11}(p)\right]\delta^{b2}
\right\}\eqno(3.4)$$
\noindent with
$$\Delta(p)=\left[1-i\frac{g}{2}L^{11}(p)\right]
            \left[1+i\frac{g}{2}L^{22}(p)\right]-
            \frac{g^2}{4}L^{12}(p)L^{21}(p).
             \eqno(3.5)$$
\noindent The propagators for physical scalar bound states
$(\bar{Q}^kQ_j) (j,k =1,\ldots,N)$ are
$$\Gamma_S^{11}(p)=i\frac{g}{2}/\left\{
                   \left[1-i\frac{g}{2}L^{11}(p)\right]-
            \frac{g^2}{4}L^{12}(p)L^{21}(p)/
            \left[1+i\frac{g}{2}L^{22}(p)\right]
            \right\}. \eqno(3.6)$$
\noindent The problem is reduced to the calculation of the fermionic loop 
$L^{ca}(p)$.  From Eqs.(3.3) and (2.8), by direct but rather lengthy 
derivation we obtain
$$L^{11}(p)=-2iI+(4m^2-p^2-i\varepsilon)\{i[K(p)+H(p)]+S(p)\}
           =[L^{22}(p)]^{*},$$
$$L^{12}(p)=L^{21}(p)=(4m^2-p^2)R(p), \eqno(3.7)$$
\noindent where $K(p), H(p), S(p)$ and $R(p)$ are all real functions and 
expressed by
$$K(p)=\frac{1}{8\pi^2}\int_{0}^{1} dx\left[
       \ln \frac{\Lambda^2+M^2(p)}{M^2(p)}-\frac{\Lambda^2}{\Lambda^2+M^2(p)}
       \right], \ \ M^2(p)=m^2-p^2x(1-x)  \eqno(3.8)$$
\noindent with the four-fermion Euclidean momentum cut-off $\Lambda$,
$$H(p)=4\pi \int\frac{d^4l}{{(2\pi)}^4}\left\{
\frac{{(l+p)}^2-m^2}{{[{(l+p)}^2-m^2]}^2+\varepsilon^2}+
(p\to -p)\right\}
\delta(l^2-m^2){\rm \sin}^2\theta(l^0,\mu), \eqno(3.9)$$
\begin{eqnarray*}
S(p)&=&4\pi^2\int \frac{d^4 l}{{(2\pi)}^4}\delta(l^2-m^2)
\delta[{(l+p)}^2-m^2] \\
& & [{\rm sin}^2\theta(l^0+p^0, \mu){\rm cos}^2\theta(l^0, \mu)+
 {\rm sin}^2\theta(l^0, \mu){\rm cos}^2\theta(l^0+p^0, \mu)] 
\ \ \ \ \ \ \ \ \ \ \ \ \ \ \ \ \ \ \ \ \ \ (3.10)
\end{eqnarray*}
\noindent and
$$R(p)=2\pi^2\int \frac{d^4 l}{{(2\pi)}^4}\delta(l^2-m^2)\delta[{(l+p)}^2-m^2]
[{\rm sin}2\theta(l^0, \mu){\rm sin}2\theta(l^0+p^0, \mu). \eqno(3.11)$$
\noindent In the above calculation, we have used the formula
$$\frac{1}{X+i\varepsilon}=\frac{X}{X^2+\varepsilon^2}-i\pi\delta(X)
\eqno(3.12)$$
\noindent and the result
$$
4\pi \int\frac{id^4l}{{(2\pi)}^4}\left\{
\frac{{[{(l-p)}^2-m^2]}^2}{{[{(l-p)}^2-m^2]}^2+\varepsilon^2}-
(p\to -p)\right\}
\delta(l^2-m^2){\rm \sin}^2\theta(l^0,\mu)$$ $$
= 4\pi\int \frac{id^4l}{{(2\pi)}^4}\left\{
[(l-p)^2-m^2]^2-(p\to -p)\right\}\pi^2 
 \delta[(l+p)^2-m^2] $$ $$
\delta(l^2-m^2)\delta[(l-p)^2-m^2]{\rm sin}^2\theta(l^0, \mu) \\
 =0\eqno(3.13)$$
\noindent owing to the fact that the arguments of the three $\delta$-functions
in the integrand can not be equal to zeros simultaneously. We notice that the
pinch singularities will appear in $S(p)$ and $R(p)$ when $p\to 0$.
Substituting Eq.(3.7) into Eq.(3.6) and taking the gap equation (2.6) into 
account, we obtain the propagator for physical scalar bound state
$({\bar{Q}}^kQ_j)$
\begin{eqnarray*}
\Gamma_S(p)&\equiv &\Gamma_S^{{(\bar{Q}^jQ_k)}^{(1)}{(\bar{Q}^kQ_j)}^{(1)}} \\
&=&-i/(p^2-4m^2+i\varepsilon)\left[
   K(p)+H(p)-iS(p)-\frac{R^2(p)}{K(p)+H(p)+iS(p)}\right]. \\
\end{eqnarray*}$$\eqno(3.14)$$
\noindent It seems that $p^2=4m^2$ is the simple pole of $\Gamma_S(p)$. 
To verify this we must examine the behavior of $K(p), H(p), S(p)$ and $R(p)$
at $p^2=4m^2$.  It is seen from Eq.(3.8) that $K(p)|_{p^2=4m^2}$ is a finite
constant when $\Lambda$ is fixed.  By means of Eq.(2.9), we may rewrite $H(p)$
in Eq.(3.9) by
\begin{eqnarray*}
H(p)&=&\frac{1}{16\pi^2|\stackrel{\rightharpoonup}{p}|}
\int_{0}^{\infty}\frac{d|\stackrel{\rightharpoonup}{l}| 
|\stackrel{\rightharpoonup}{l}|}{\omega_l}\left[
\ln\frac{(p^2-2\omega_lp^0+2|\stackrel{\rightharpoonup}{l}|
|\stackrel{\rightharpoonup}{p}|)^2+\varepsilon^2}
        {(p^2-2\omega_lp^0-2|\stackrel{\rightharpoonup}{l}|
|\stackrel{\rightharpoonup}{p}|)^2+\varepsilon^2}
   +(p^0\to -p^0)\right] \\
& & \left\{1/[e^{\beta(\omega_l-\mu)}+1]+
                               1/[e^{\beta(\omega_l+\mu)}+1]
                        \right\}, \ \ \ 
\omega_l=\sqrt{{\stackrel{\rightharpoonup}{l}}^2+m^2},
\end{eqnarray*}               
$$\eqno(3.15)$$
\noindent where the zero points of the arguments of all the logrithmic
function must be removed from the integral because these functions come from 
the integration of the principal parts of the integrand. It is indicated that
$H(p)$ in Eq.(3.15) contains no singularity when 
$|\stackrel{\rightharpoonup}{p}| \to 0$.  In fact, if we set $p^2=\lambda^2$,
then when $|\stackrel{\rightharpoonup}{p}| \to 0, \ p^0=\lambda$ and it can
be proven that
$$\lim \limits_{|\stackrel{\rightharpoonup}{p}| \to 0}
\frac{1}{|\stackrel{\rightharpoonup}{p}|}  
\ln\frac{(p^2\mp 2\omega_lp^0+2|\stackrel{\rightharpoonup}{l}|
|\stackrel{\rightharpoonup}{p}|)^2+\varepsilon^2}
        {(p^2\mp 2\omega_lp^0-2|\stackrel{\rightharpoonup}{l}|
|\stackrel{\rightharpoonup}{p}|)^2+\varepsilon^2}=
\frac{8|\stackrel{\rightharpoonup}{l}|(\lambda^2\mp 2\omega_l\lambda)}
    {(\lambda^2\mp 2\omega_l\lambda)^2+\varepsilon^2}
\eqno(3.16)$$
\noindent which are finite even if when $\lambda=0$.  It is easy to find
that when $p^2=4 m^2$ the arguments of the logrithmic functions in Eq. (3.15)
have no zero except $p^2-2\omega_lp^0+2|\stackrel{\rightharpoonup}{l}|
|\stackrel{\rightharpoonup}{p}|=0$ if $|\stackrel{\rightharpoonup}{l}|=
|\stackrel{\rightharpoonup}{p}|/2$.  However, now that this point has been
removed from the integral, it will not lead any singularity of $H(p)$.\\
\indent When $p^2=4 m^2$, the general form of $S(p)$ and $R(p)$ may be 
expressed by
\begin{eqnarray*}
A(p)|_{p^2=4 m^2}&=&\int d^4l\delta(l^2-m^2)\delta[(l+p)^2-m^2]
             f(l^0,p^0,\mu)|_{p^2=4m^2} \\
                 &=&\int \frac{d^3l}{4\omega_l}\left[
\delta(\omega_lp^0-
|\stackrel{\rightharpoonup}{l}||\stackrel{\rightharpoonup}{p}|{\rm cos}\theta+
2m^2)f(\omega_l,p^0,\mu) \right. \\
& & \left.\left. +\delta(-\omega_lp^0-
|\stackrel{\rightharpoonup}{l}||\stackrel{\rightharpoonup}{p}|{\rm cos}\theta+
2m^2)f(-\omega_l,p^0,\mu)
\right]\right|_{p^0=\sqrt{{\stackrel{\rightharpoonup}{p}}^2+4m^2}}.
\end{eqnarray*}
$$\eqno(3.17)$$
\noindent Since $|{\rm cos}\theta|\leq 1$, the argument of the first 
$\delta$-function in Eq.(3.17) could not be zero for any value of 
$|\stackrel{\rightharpoonup}{l}|$ and the second $\delta$-function could have
zero argument only if $|\stackrel{\rightharpoonup}{l}|=
|\stackrel{\rightharpoonup}{p}|/2 \ ({\rm cos}\theta=-1)$, thus we obtain
$$A(p)|_{p^2=4 m^2}=\left.\pi \int_{0}^{\infty}
\frac{d |\stackrel{\rightharpoonup}{l}|}{4\omega_l}
f(-\omega_l,p^0,\mu)\delta_{|\stackrel{\rightharpoonup}{l}|,
\frac{|\stackrel{\rightharpoonup}{p}|}{2}}
\right |_{p^0=\sqrt{{\stackrel{\rightharpoonup}{p}}^2+4m^2}}=0.
\eqno(3.18)$$
\noindent This means that
$$S(p)|_{p^2=4m^2}=R(p)|_{p^2=4m^2}=0.
\eqno(3.19)$$
\noindent As a result, the propagator of the scalar bound state 
$(\bar{Q}^kQ_j)$
$$\Gamma_S(p) \rightarrow  
-i/(p^2-4m^2+i\varepsilon)[K(p)+H(p)], \ \ \ \ {\rm when} \ \ p^2\to 4m^2
\eqno(3.20)$$
\noindent and $p^2=4m^2$ is its simple pole indeed.  In this way, we obtain
$N^2$ scalar bound states $(\bar{Q}^kQ_j) \ (j,k=1,...,N)$ with the mass 
$2m$. \\
\noindent It may be verified that $\Gamma_S(p)$ expressed by Eq.(3.14)
contains no pinch singularity.  We notice that when $p \to 0$ $K(p)$ is still
a finite constant, and $H(p)=0$ by Eq.(3.9) and $S(p)-R(p)=0$ from Eqs.(3.10)
and (3.11).  These results indicate that the terms containing pinch singularity
in the denominator of $\Gamma_S(p)$ will become
$$-iS(p)-\frac{R^2(p)}{K(p)+iS(p)}
\stackrel{p\to 0}{\longrightarrow}\frac{-iS(p)K(p)+S^2(p)-R^2(p)}{S^2(p)}
\rightarrow 0,
\eqno(3.21)$$
\noindent i.e. the pinch singularities in the propagator $\Gamma_S(p)$ will 
be cancelled by each other and do not appear in the final expression.
\section{Pseudoscalar bound state modes}  
A pararell discussion to scalar bound states can be applied to the case with
pseudoscalar bound states.  The relevant four-fermion interactions are now 
expressed by the Lagrangian
$${\cal L}_{int}^P=\frac{g}{4}\sum_{a=1,2}\sum_{j,k=1}^{N}
{(-1)}^{a+1}{(\bar{Q}^ji\gamma_5Q_k)}^{(a)}{(\bar{Q}^ki\gamma_5Q_j)}^{(a)}.
\eqno(4.1)$$
\noindent For pseudoscalar bound state $(\bar{Q}^ki\gamma_5Q_j)$, the 
corresponding pseudoscalar four-point function from $a$-type vertex to 
$b$-type vertex can be denoted by
$$\Gamma_P^{{(\bar{Q}^ji\gamma_5Q_k)}^{(b)}{(\bar{Q}^ki\gamma_5Q_j)}^{(a)}}(p)
\equiv \Gamma_P^{ba}(p)$$
\noindent and submit to the algebraic equations
$$\Gamma_P^{ba}(p)=i\frac{g}{2}{(-1)}^{a+1}\delta^{ba}+
\sum_{c=1,2}\Gamma_P^{bc}(p)N^{ca}(p)i\frac{g}{2}{(-1)}^{a+1}, \ \ a,b=1,2
\eqno(4.2)$$
\noindent where $N^{ca}(p)$ expresses the contribution of the $Q_j-\bar{Q}^k$
fermion loop with an $a$-type and a $c$-type pseudoscalar interaction vertex,
i.e.
$$N^{ca}(p)\equiv N_{\bar{Q}^kQ_j}^{ca}(p)=
-\int \frac{d^4l}{{(2\pi)}^4}tr[i\gamma_5iS^{ca}(l,m)i\gamma_5iS^{ac}(l+p,m)].
\eqno(4.3)$$
\noindent It is easy to see that Eq.(4.2) has the same form as Eq.(3.2)
after the substitutions $\Gamma_P^{ba}(p)\to \Gamma_S^{ba}(p)$ and
$N^{ca}(p)\to L^{ca}(p)$.  Hence we can directly put down the propagators for
physical pseudoscalar bound states $(\bar{Q}^ki\gamma_5Q_j) \ (j,k=1,...,N)$
$$\Gamma_P(p)\equiv \Gamma_P^{11}(p)=i\frac{g}{2}/\left\{
                   \left[1-i\frac{g}{2}N^{11}(p)\right]-
            \frac{g^2}{4}N^{12}(p)N^{21}(p)/
            \left[1+i\frac{g}{2}N^{22}(p)\right]
            \right\}. \eqno(4.4)$$
\noindent The results of calculations of $N^{ca}(p)$ are
$$N^{11}(p)=-2iI-i(p^2+i\varepsilon)[K(p)+H(p)]-iS(p)]
           =[N^{22}(p)]^{*},$$
$$N^{12}(p)=N^{21}(p)=-p^2R(p). \eqno(4.5)$$
\noindent Thus we obtain
$$\Gamma_P(p)
=-i/(p^2+i\varepsilon)\left[
   K(p)+H(p)-iS(p)-\frac{R^2(p)}{K(p)+H(p)+iS(p)}\right].
\eqno(4.6)$$
\noindent We observe that when $p^2\to 0$, $K(p)$ is finite, $H(p)$ in 
Eq.(3.9) is equal to zero and both $S(p)$ and $R(p)$ are also equal to zeroes
because the arguments of $\delta(l^2-m^2)$ and $\delta[(l+p)^2-m^2]$ in 
Eqs.(3.10) and (3.11) can not be zeroes simultaneously.  Consequently, we have
$$\Gamma_P(p) \stackrel{p^2\to 0}{\longrightarrow}  
-i/(p^2+i\varepsilon)K(p)
\eqno(4.7)$$
\noindent which is of the same form as the propagator for pseudoscalar bound
state at $T=0$.  Therefore, $p^2=0$ is the simple pole of 
$\Gamma_P(p)$ and we will have $N^2$ massless pseudoscalar bound
states $(\bar{Q}^ki\gamma_5Q_j) \ (j,k=1,...,N)$. By comparing Eq.(4.6) with 
Eq.(3.14) we see that $\Gamma_P(p)$ and $\Gamma_S(p)$ have the identical
form except the position of the pole. Hence the same cancellation mechanism
of the pinch singularities as in  $\Gamma_S(p)$ certainly exists in 
$\Gamma_P(p)$ as well and we need not worry about the problem of pinch 
singularity here.
\section{Conclusion}
The above discussions show that under the assumption (2.5) i.e. all the
fermions have equal masses and equal chemical potentials,
at the finite temperature $T<T_c$, the
critical temperature below which the gap equation (2.6) is satisfied,
 we may obtain $N^2$ scalar bound states 
$(\bar{Q}^kQ_j)\ (j,k=1,...,N)$ with the mass $2m$ and $N^2$ massless 
pseudoscalar bound states $(\bar{Q}^ki\gamma_5Q_j)\ (j,k=1,...,N)$.
These results characterize spontaneous symmetry breaking of the chiral group 
$U_L(N)\times U_R(N)$ down to the vector-like group $U_{L+R}(N)$. 
The $N^2$ massive scalar composite particles will correspond to the generators
of the unbroken group $U_{L+R}(N)$.  The $N^2$ massless pseudoscalar 
composite particles will correspond to the generators of the broken 
axial-vector group $U_{L-R}(N)$ and can be identified with the Nambu-Goldstone
bosons.  This shows  the Goldstone Theorem  at finite temperature.  Here the
theorem is proven in the chiral $U_L(N)\times U_R(N)$ model of NJL-form by
means of the real-time formalism of thermal field theory without any
incosistency.  However, we emphesize that the assumption (2.5) is decisive for
validity of such consistency between the Goldstone Theorem at finite temperature
and the real-time thermal field theory. For the model discussed in this paper,
the assumption (2.5), especially that the fermions have the same masses, can be
natural and plausible.  As for the models in which the assumption (2.5) could
not satisfied, we will research them elsewhere.\\ \\

\end{document}